**Estimating breast cancer recurrence in a population-based registry in Georgia, US**

Chrystelle Kiang,[1] Micah Streiff,[1] Rebecca Nash,[1] Robert Lyles,[2] Deirdre Cronin-Fenton,[3] Anke Huels,[2] Timothy L. Lash,[1] Kevin C. Ward[1]

[1] Department of Epidemiology, Rollins School of Public Health, Emory University
[2] Department of Biostatistics and Bioinformatics, Rollins School of Public Health, Emory University
[3] Department of Clinical Epidemiology, Aarhus University Aarhus University and Hospital

**ABSTRACT**

Although the descriptive epidemiology of primary breast cancer is well characterized in the US, breast cancer recurrence rates have not been measured in an unselected population. The number of breast cancer survivors at risk for recurrence is growing each year, so recurrence surveillance is a pressing need**.** We used missing data methods to impute breast cancer recurrence and estimate the risk of recurrence in the Cancer Recurrence Information and Surveillance Program (CRISP) cohort in the Georgia Cancer Registry. The imputation model was based on an internal validation substudy and indicators recorded in the registry (e.g., pathology reports, imaging claims), prognostic variables (e.g., stage at diagnosis), and characteristics associated with missing data (e.g., insurance coverage). We pooled hazard ratios (HR) and 95% Confidence Intervals (CI) across 1000 imputed datasets, adjusted for age, stage, grade, subtype, race and ethnicity, marital status, and urban/rural county at diagnosis. There were 1,606 patients with a validated outcome (75% with breast cancer recurrence) and we imputed the outcome for the remaining 23,439 patients. We estimated an overall 7.2% incidence of recurrence between at least 1 year after diagnosis and up to 5 years of follow up. When comparing the hazards pooled across imputations, we found that some patterns differed from established patterns in mortality or survival, notably by race and ethnicity, underscoring the need for continued research on the descriptive epidemiology of breast cancer recurrence. These results provide new insights into surveillance for breast cancer survivors in Georgia, especially those with higher stage and grade tumors, of Hispanic ethnicity, and who may be lacking social support.

**INTRODUCTION**

The epidemiology of primary breast cancer is well characterized in the US in part due to the long-standing history of population-based cancer registries, but the descriptive epidemiology of recurrence among breast cancer survivors has not been reliably captured. The number of female breast cancer survivors at risk for recurrence is growing, thanks to improvements in cancer screening and treatment, and as deaths due to infectious and cardiovascular disease decrease.[1–3] Thus, there is a need to estimate the incidence of recurrence and profile the recurrent population.[4,5]

Surveillance systems were originally designed for counting new cancer primaries, adding the systematic collection of recurrence data in the US must essentially be retrofit to the existing surveillance mechanisms. Challenges of recurrence surveillance in existing systems include the complexity of patient trajectories, the mobility of patients over time, and the overlap in signals for ongoing primary treatment, subsequent primaries, and recurrence. Additionally, there is often variability in the available data by health insurance providers, where more information may be more readily shared for patients with insurance through the government programs Medicare and Medicaid through linkage compared to patients with private insurance. There has been one approach with National Program of Cancer Registries (NPCR) registries across 5 states by Thompson et al., in which they collected medical records from all healthcare facilities for up to 5 years for 11,769 breast cancer patients diagnosed in 2011 and estimated 6% of patients experienced a recurrence during the follow-up period for all stages.[6] However, their approach was resource intensive and relied heavily on abstractors following all patients, highlighting the need for a more efficient and sustainable approach.

In the state of Georgia, cancer surveillance is operated by the population-based Georgia Cancer Registry (GCR), which receives information about diagnoses, cancer-directed medical procedures and treatments for residents of the state. To capture recurrence systematically, the GCR launched the Cancer Recurrence Information and Surveillance Program (CRISP) to surveil recurrence for four cancers, with breast cancer as the initial focus. The study design was to use multiple data streams that would provide signals of potential recurrence, use medical records to validate those signals in subsets of patients (i.e., a validation cohort), and then refine the signals based on the validation work. Oncology Data Specialists reviewed the medical records of those in the validation cohort to determine whether they had truly experienced a recurrence. The goal of this paper is to estimate the incidence of and provide novel descriptive epidemiology of breast cancer recurrence in the CRISP cohort. We used the results of the validation cohort to inform preliminary results among the larger cohort, for which the signals but not the medical record validation were available.

**METHODS**

**Study Population**

The study population consisted of female breast cancer patients in the GCR who were diagnosed between 2013 and 2017 with Stage I, II, or III disease. The start of follow up for recurrences began at a derived disease-free date defined as one year after diagnosis, or if there was ongoing treatment in a year, as 60 days after the last record of treatment. Patients were excluded if they experienced a recurrence or second primary cancer, emigrated from the state, or died before their disease-free date. Patients were also excluded if they did not receive any first course treatment or did not reach disease-free status before end of the study on December 31, 2019 (Figure 1).

**Data Sources**

The data routinely sent to GCR include pathology reports that contain cancer terminology, hospital data from CoC-accredited facilities, hospital discharge records, and cause of death from the National Death Index. The novel data streams that were linked to the cancer registry as part of CRISP included claims data reported by facilities that provide cancer services such as hospitals, outpatient oncology and/or surgery facilities, laboratories, pharmacies, and private physicians and physician's offices, as well as Medicare and Medicaid data.[7] Treatment information was collected from these sources. Emigration data are provided by LexisNexis and death data are from state vital records, the National Death Index, and hospice facilities.

**Study Variables**

<u>Outcome</u>

The outcome of interest was breast cancer recurrence, defined as clinical evidence of the original cancer in any organ site following definitive treatment and a disease-free period. This definition did not include a second primary diagnosis in the same or contralateral breast cancer unless specifically confirmed through pathology to be a recurrence of the original primary. For the validation cohort, the observed (true) outcome was determined by a review of medical records by tumor registrars and the outcome was missing for the remaining cohort.

<u>Covariates</u>

We also included variables related to prognostic factors of tumor stage, tumor grade, molecular subtype, age at diagnosis, and demographic variables of race and Hispanic ethnicity, marital status, and urbanicity of county of residence, as reported in GCR at time of diagnosis and categorized according to SEER standards. However, because of small numbers and to protect patient privacy, for race and Hispanic ethnicity, non-Hispanic (NH) patients who were American Indian/Alaskan Native, other race, or unknown to patient were combined into one NH Other category. We included death and emigration data for follow up. We also included Medicare/Medicaid coverage and Health Maintenance Organization (HMO) coverage among those with Medicare, each a binary indicator of whether patients had at least one month of coverage during the study period.

Recurrence Indicators

We derived a set of nine indicators as potential signals of recurrence for the entire cohort based on prior registry knowledge of primary treatment and how diagnoses of recurrence would appear in records. The nine dichotomous indicators are: validated record of pathology reported to GCR after disease-free date (independent of secondary malignant neoplasm (SMN) diagnosis codes or systemic therapy); pathology reported to GCR & recent SMN diagnosis; pathology reported to GCR & recent systemic therapy; pathology in medical claims data & recent SMN diagnosis; pathology in medical claims data & recent systemic therapy; imaging record & recent SMN; imaging record & recent therapy; death & recent SMN diagnosis; CoC determined recurrence. These indicators are combinations of variables from both routine reporting to GCR and new linkage as part of CRISP. From the routine data, there was a pathology variable if there was any pathology report identified by existing algorithms as being cancer-related and variables for death or a CoC-identified recurrence. From the new linkage to medical claims data, there was a separate pathology variable and variables if patients had record of imaging reports, secondary malignant neoplasm (SMN) diagnoses, or new systemic therapy. The systemic therapies are treatments that may have been used to target for recurrence: chemotherapy, radiation, immunotherapy. These are usually part of primary treatment and do not include maintenance therapies such as endocrine therapy. The pathology and imaging reports reflect whether the record exists but not the result, as the claims data only contain information that the test was completed and cancer-related. We combined SMN and systemic therapy with pathology and imaging if either was done in the 30 days before or 60 days after the pathology or imaging signal, and the appropriate combination would need to be fulfilled to be considered as a positive indicator. If patients had multiple of the same indicator, it was counted only at the first occurrence.

**Statistical Analyses**

First, we summarized the descriptive statistics of the study population, grouped by membership into the validation cohort, tabulating count and percentage of demographic and clinical characteristics. To estimate recurrence among the main cohort, for whom the outcome was not observed or ascertained because they were not in the validation cohort, we used multiple imputation, specifically using multivariate imputation by chained equations (MICE) with the *mice* package in R.[8] We framed our outcome of recurrence as a missing data problem, where we have the observed data for the validation cohort, while the remainder of the cohort is the unobserved group where data are missing at random. Selection into the validation cohort (i.e., missingness) was not completely at random and was dependent on patients having procedures done in response to suspected recurrence and having the accompanying records sent to the registry. Records were more likely to be sent to the registry for patients with health insurance coverage (Medicare/Medicaid, HMO) and we have health insurance status at diagnosis for patients, which we included in our model as missing data indicators. We also included year of diagnosis to account for differences in time to contribute data.

For those with missing data, we imputed a binary outcome of recurrence based on the nine indicators previously listed, tumor stage, tumor grade, tumor subtype, age at diagnosis, death (any cancer and all-cause), race and ethnicity, marital status, urbanicity of the county at diagnosis, year of diagnosis, HMO coverage, and Medicare coverage. When the outcome was imputed as a recurrence, an accompanying continuous time to recurrence was also imputed based on the same variables. If the outcome was imputed as not a recurrence, the observed censor time was used for follow-up time. Those who were imputed with a recurrence in at least half of all imputations were classified as recurrent cases and along with those with validated recurrences, comprised the recurrent cohort. We summarized the descriptive statistics of the recurrent cohort and presented the mean and median imputed time to recurrence across imputations. We estimated adjusted pooled hazard ratios (HRs) and accompanying 95% CI of recurrence, where the regression was applied to each imputed dataset, then pooled together to produce adjusted measures of association and confidence intervals that incorporate the variation between imputations. All analyses were completed using R Statistical Software (v4.3.0).[9] Though the study period is through 2019, there is ongoing work in validating records by the GCR and these analyses are based on data updated through June 2025.

## RESULTS

Of the 27,453 eligible patients, the study population consisted of 25,045 patients, with 1,606 in the validation cohort and 23,439 in the main cohort with an unobserved outcome. The descriptive characteristics of the study population are in Table 1. Patients in the validation cohort more often had later stage, higher grade, and TNBC subtype tumors than the main cohort. The validation cohort included higher prevalence of patients of NH Black and Hispanic race, and similar distribution by marital status and county. Most of both cohorts did not have HMO coverage and about half of both cohorts had Medicare coverage. By the end of the study period of December 31, 2019, 56% of the validation cohort and 94% of the main cohort were alive. There were a total of 1,814 (7.2%) recurrent cases identified across the validation and main cohorts. Among the validation cohort, 1,201 of 1,606 (74.8%) had a recurrence validated by medical record review, while 613 of 23,439 (2.6%) of the main cohort were imputed as having a recurrence in at least half of imputations. Figure 2 shows a histogram of the count of individuals in the unobserved cohort and their mean recurrence, or probability of being imputed as recurrence, across imputations. For the recurrent cases, the mean and median (IQR) follow up time after disease-free date was 21.4 and 20.7 (16.8, 25.1) months. For those who did not have a recurrence, the mean and median (IQR) follow up time after disease-free date was 37.3 and 38.6 (38.3, 38.6) months. Among those identified or imputed with a recurrence, most were between ages of 45 and 64, had Stage II, grade 3 or higher, and Luminal A subtype tumor, were of NH White race, married at diagnosis, and living in an urban county (Table 2).

The HRs for the outcome of recurrence adjusted for age, grade, stage, subtype, race/ethnicity, marital status, and county, were pooled across imputations and shown in

Table 3. Compared to the referent group of age 55–64 at diagnosis, those aged under 35 had decreased hazard of recurrence (HR for ages 18–34 compared with 55–64 was 0.74; 95% CI 0.45, 1.23). For other age groups, there was slight decrease for ages 35–44 with HR (95% CI) of 0.94 (0.65, 1.36) and 0.93 (0.64, 1.34) for ages 65–74. Those aged 35–44 and over 75 had slightly increased hazard with HR (95% CI) of 1.19 (0.77, 1.82) for ages 34–44 and 1.32 (0.84, 2.08) for ages 75 and up. For both tumor stage and grade, the hazard increased as stage and grade increased. There was no notable difference by tumor subtype or county of residence. Compared to referent of NH White patients, Hispanic and NH API patients had slightly elevated hazard of recurrence with HR (95% CI) of 1.35 (0.75, 2.44) ) for Hispanic patients and 1.03 (0.45, 2.39) for NH API patients. NH Black patients had slightly decreased hazard with HR (95% CI) 0.94 (0.73, 1.21). Compared to married patients, there was increased hazard for divorced or widowed patients, with HR (95%) of 1.53 (1.04, 2.26) and 1.23 (0.81, 1.87), respectively. There was equal hazard for single/never married patients 1.00 (0.72, 1.39).

## DISCUSSION

Among a cohort of breast cancer patients, stage I–III, who completed primary treatment in Georgia, US, we estimated an overall 7% incidence of breast cancer recurrence over 5 years. In our adjusted models pooled across imputations, the rate of recurrence increased as stage and grade increased, was greater for those aged 35 to 64 than those under 35 and over 65 but did not differ much by tumor subtype. The rate among NH Black patients was similar to the rate among NH White patients, while the rate was slightly increased for Hispanic and NH API patients. Patients who were widowed had an increased rate of recurrence, even after adjusting for age (at diagnosis), while the rate was similar between those in an urban or rural county. Our estimate of recurrence incidence is consistent with previous literature. In the NPCR study by Thompson et al. using US-based registry data, they identified 8,670 breast cancer cases at stage I–III, and were able to establish a disease-free date for 8,033 patients, of whom 575 (7%) had a recurrence over five years of follow-up.[6].

Other attempts to capture breast cancer recurrence in SEER registries involve developing algorithms to predict recurrent cases based on the data structure among the study sample and do not provide the same scope of estimates as our work. Ritzwoller et al. developed an algorithm for breast cancer recurrence that used some similar potential indicators of recurrence as ours, as well as some data that would only be available within HMOs and not necessarily provided to GCR, so may not be scalable in a population-based registry.[10] Their approach also requires a larger proportion of validated data to inform their model. They applied their algorithm to two test datasets and for one among 3,370 patients, estimated 7.2% recurrences over a median of 53 months of follow-up (range 1–156) and among 3,961 patients, 7.6% over a median of 75 months of follow up (range 1–112).[10] These are similar to our estimate of 7.2% but we had shorter follow up, with a median 37 of months. In another approach, A'mar et al. used a data mining algorithm to identify recurrences among 3,152 patients and estimated 25% recurrence risk over 14 years of follow-up.[11] This duration was much longer than our follow up and is dependent on having complete medical records, which is not applicable in the context of a US population-based registry.[11]

Our results provide insight on breast cancer outcomes to complement existing literature about mortality. We found that there was a higher rate of recurrence for patients who were no longer married (divorced or separated) or widowed compared to other marital statuses, which is consistent with findings of worse survival, perhaps suggesting that a sudden loss of social support (at likely older age for widows) means worse prognosis.[12] We did not observe this difference in rates between those who were never married and single or currently married. Our findings among groups of race and ethnicity differed from published mortality outcomes. NH Black and NH White patients had similar rate of recurrence, once adjusted for age, stage, grade, subtype, yet there is evidence that NH Black women are disproportionately affected by cancer mortality.[13,14] We found that compared to NH White patients, there was slightly increased rate of recurrence among Hispanic patients; though this difference is not observed in breast cancer mortality. This pattern may be partially explained by the timing of recurrence by race and ethnicity, especially those before our derived disease-free date. Among those selected for validation, there was a larger proportion of NH Black and Hispanic patients who experienced a recurrence before our derived disease-free date, but the validation also oversampled NH Black patients and patients of later stage compared to overall population. These differences in recurrence rate, overall and even by year, and mortality patterns highlight the need for recurrence surveillance and merits additional research.

Strengths and Limitations

This study was the first to estimate the incidence of recurrence at the population-level in the US, using existing registry data. The population of the state of Georgia is large and diverse enough to allow us to compare recurrence by groups of race and ethnicity, which is important to capture given the differences in the prevalence of luminal subtypes and the disparities in breast cancer mortality outcomes. In using multiple imputation, we were able to optimize the use of rich data available in the registry and to incorporate all data from the validation cohort. Not all combinations of indicators occurred in both the validation and main cohorts. This lack of overlap could make traditional methods more difficult, but using multiple imputation allowed for the borrowing of information from other variables without requiring an exact match of covariate patterns. The validation cohort was not representative of the main cohort nor a completely random sample, but neither are requirements for using multiple imputation. There is no way to quantify the mechanism of missing data, but our missing data are dependent on factors largely known to us and that have observed data for all members of the cohort, which is consistent with descriptions of missing at random.[19–21] We made best efforts with the data we had to ensure the model was appropriate, including variables related to selection into the validation cohort.[22] We used a large number of imputations to allow for more robust estimation given the large population size and presented pooled results to reflect the uncertainty. Still, we may not have fully captured the probability of missingness, and these results are dependent on our assumptions about the missing data mechanism.

There is a possibility for information bias as the availability of different types of data differed for patients, and indicators and disease-free dates were not always ascertained in the same way. That knowledge was incorporated into the definition of indicators and disease-free dates. Capturing recurrence remains a challenge even in places with government-sponsored healthcare and cohesive health data systems. This challenge has been addressed in Denmark and Sweden, where algorithms have been developed to capture recurrent cases in medical records or administrative claims data, although even these algorithms are imperfect.[15–17] A validation study of the capture in Sweden, for example, found that recurrence capture was not as good as capture of primary diagnosis.[18]

We addressed the potential of immortal person time where patients may not yet be at risk of recurrence, we derived a conservative estimate of disease-free date. In the NCPR work by Thompson et al., they found the median time from surgery and disease-free status was 206, 147, 193, and 185 days for four of the states included.[6] These were all below our minimum of 365 days, providing reassurance about possible bias from including immortal person time, however, our result excludes very early recurrences, which may differ in rate between patients and overall may be an underestimation of recurrence.

**CONCLUSION**

In conclusion, we estimated the overall incidence of breast cancer recurrence to be ~7% for female breast cancer patients in Georgia, in the period at least 1 year after diagnosis after first course of treatment with curative intent with up to 5 years of follow up. We provided the descriptive epidemiology of the recurrence cases and described relative rates by prognostic variables and demographic characteristics using estimates pooled across imputations, specifically noting that the rate of recurrence increased for those diagnosed with later grade and stage, for Hispanic patients compared to NH White patients, and those who were no longer married (separated, divorced, widowed) compared to those married. The approach of using signals from registry data streams with multiple imputation methods to estimate the recurrence can be used to further characterize the descriptive epidemiology of breast cancer for longer follow up time, as well as for other types of cancers.

# REFERENCES

1. Brewster AM, Hortobagyi GN, Broglio KR, Kau SW, Santa-Maria CA, Arun B, Buzdar AU, Booser DJ, Valero V, Bondy M, Esteva FJ. Residual Risk of Breast Cancer Recurrence 5 Years After Adjuvant Therapy. *J Natl Cancer Inst*. 2008;100(16):1179-1183. doi:10.1093/jnci/djn233
2. Bosco JLF, Lash TL, Prout MN, Buist DSM, Geiger AM, Haque R, Wei F, Silliman RA. Breast cancer recurrence in older women five to ten years after diagnosis. *Cancer Epidemiol Biomarkers Prev*. 2009;18(11):2979-2983. doi:10.1158/1055-9965.EPI-09-0607
3. Pedersen RN, Esen BÖ, Mellemkjær L, Christiansen P, Ejlertsen B, Lash TL, Nørgaard M, Cronin-Fenton D. The Incidence of Breast Cancer Recurrence 10-32 Years after Primary Diagnosis. *J Natl Cancer Inst*. Published online November 8, 2021:djab202. doi:10.1093/jnci/djab202
4. Warren JL, Yabroff KR. Challenges and Opportunities in Measuring Cancer Recurrence in the United States. *J Natl Cancer Inst*. 2015;107(8):djv134. doi:10.1093/jnci/djv134
5. In H, Bilimoria KY, Stewart AK, Wroblewski KE, Posner MC, Talamonti MS, Winchester DP. Cancer Recurrence: An Important but Missing Variable in National Cancer Registries. *Ann Surg Oncol*. 2014;21(5):1520-1529. doi:10.1245/s10434-014-3516-x
6. Thompson TD, Pollack LA, Johnson CJ, Wu XC, Rees JR, Hsieh MC, Rycroft R, Culp M, Wilson R, Wu M, Zhang K, Benard V. Breast and colorectal cancer recurrence and progression captured by five U.S. population-based registries: Findings from National Program of Cancer Registries patient-centered outcome research. *Cancer epidemiology*. 2020;64:101653. doi:10.1016/j.canep.2019.101653
7. Georgia Comprehensive Cancer Registry | Georgia Department of Public Health. Accessed July 3, 2024. https://dph.georgia.gov/chronic-disease-prevention/georgia-comprehensive-cancer-registry
8. Buuren S van, Groothuis-Oudshoorn K. mice: Multivariate Imputation by Chained Equations in R. *Journal of Statistical Software*. 2011;45:1-67. doi:10.18637/jss.v045.i03
9. R Core Team. R: A Language and Environment for Statistical Computing. Published online 2024. https://www.R-project.org
10. Ritzwoller DP, Hassett MJ, Uno H, Cronin AM, Carroll NM, Hornbrook MC, Kushi LC. Development, Validation, and Dissemination of a Breast Cancer Recurrence Detection and Timing Informatics Algorithm. *JNCI Journal of the National Cancer Institute*. 2017;110(3):273. doi:10.1093/jnci/djx200
11. A'mar T, Beatty JD, Fedorenko C, Markowitz D, Corey T, Lange J, Schwartz SM, Huang B, Chubak J, Etzioni R. Incorporating Breast Cancer Recurrence Events Into Population-Based Cancer Registries Using Medical Claims: Cohort Study. *JMIR Cancer*. 2020;6(2):e18143. doi:10.2196/18143
12. Hjorth CF, Damkier P, Ejlertsen B, Lash T, Sørensen HT, Cronin-Fenton D. Socioeconomic position and prognosis in premenopausal breast cancer: a population-based cohort study in Denmark. *BMC Med*. 2021;19:235. doi:10.1186/s12916-021-02108-z
13. Collin LJ, Jiang R, Ward KC, Gogineni K, Subhedar PD, Sherman ME, Gaudet MM, Breitkopf CR, D'Angelo O, Gabram-Mendola S, Aneja R, Gaglioti AH, McCullough LE. Racial Disparities in Breast Cancer Outcomes in the Metropolitan Atlanta Area: New Insights and Approaches for Health Equity. *JNCI Cancer Spectr*. 2019;3(3):pkz053. doi:10.1093/jncics/pkz053

14. Moubadder L, Collin LJ, Nash R, Switchenko JM, Miller-Kleinhenz JM, Gogineni K, Ward KC, McCullough LE. Drivers of racial, regional, and socioeconomic disparities in late-stage breast cancer mortality. *Cancer*. 2022;128(18):3370-3382. doi:10.1002/cncr.34391
15. Cronin-Fenton D, Kjærsgaard A, Nørgaard M, Amelio J, Liede A, Hernandez RK, Sørensen HT. Breast cancer recurrence, bone metastases, and visceral metastases in women with stage II and III breast cancer in Denmark. *Breast Cancer Res Treat*. 2018;167(2):517-528. doi:10.1007/s10549-017-4510-3
16. Jensen AR, Storm HH, Møller S, Overgaard J. Validity and Representativity in the Danish Breast Cancer Cooperative Group. *Acta Oncologica*. 2003;42(3):179-185. doi:10.1080/02841860310000737
17. Wadsten C, Heyman H, Holmqvist M, Ahlgren J, Lambe M, Sund M, Wärnberg F. A validation of DCIS registration in a population-based breast cancer quality register and a study of treatment and prognosis for DCIS during 20 years: Two decades of DCIS in Sweden. *Acta Oncologica*. 2016;55(11):1338-1343. doi:10.1080/0284186X.2016.1211317
18. Palmér S, Blomqvist C, Holmqvist M, Lindman H, Lambe M, Ahlgren J. Validation of primary and outcome data quality in a Swedish population-based breast cancer quality registry. *BMC Cancer*. 2024;24:329. doi:10.1186/s12885-024-12073-4
19. Van Buuren S. *Flexible Imputation of Missing Data*. CRC press; 2018.
20. van Ginkel JR, Linting ,Marielle, Rippe ,Ralph C. A., and van der Voort A. Rebutting Existing Misconceptions About Multiple Imputation as a Method for Handling Missing Data. *Journal of Personality Assessment*. 2020;102(3):297-308. doi:10.1080/00223891.2018.1530680
21. Higgins JP, White IR, Wood AM. Imputation methods for missing outcome data in meta-analysis of clinical trials. *Clin Trials*. 2008;5(3):225-239. doi:10.1177/1740774508091600
22. Madley-Dowd P, Hughes R, Tilling K, Heron J. The proportion of missing data should not be used to guide decisions on multiple imputation. *Journal of Clinical Epidemiology*. 2019;110:63-73. doi:10.1016/j.jclinepi.2019.02.016

# TABLES AND FIGURES

Table 1. Patient and tumor characteristics by cohort membership of female stage I–III breast cancer patients in the Georgia Cancer Registry, diagnosed 2013 to 2017 with follow-up through 2019.

| Characteristic | Validation cohort N = 1,606 | Main cohort N = 23,439 |
|---|---|---|
| Diagnosis Year | | |
|   2013 | 431 (27%) | 4,295 (18%) |
|   2014 | 376 (23%) | 4,555 (19%) |
|   2015 | 338 (21%) | 4,620 (20%) |
|   2016 | 288 (18%) | 4,909 (21%) |
|   2017 | 173 (11%) | 5,060 (22%) |
| Age at diagnosis | | |
|   18-34 | 63 (3.9%) | 522 (2.2%) |
|   35-44 | 224 (14%) | 2,470 (11%) |
|   45-54 | 362 (23%) | 5,179 (22%) |
|   55-64 | 362 (23%) | 6,519 (28%) |
|   65-74 | 401 (25%) | 5,868 (25%) |
|   75-100 | 194 (12%) | 2,881 (12%) |
| Stage | | |
|   I | 421 (26%) | 12,643 (54%) |
|   II | 708 (44%) | 8,522 (36%) |
|   III | 477 (30%) | 2,274 (9.7%) |
| Grade | | |
|   1 | 134 (8.3%) | 5,276 (23%) |
|   2 | 591 (37%) | 9,932 (42%) |
|   3+ | 840 (52%) | 7,477 (32%) |
|   Unknown | 41 (2.6%) | 754 (3.2%) |
| Subtype | | |
|   Luminal A | 920 (57%) | 16,062 (69%) |
|   Luminal B | 195 (12%) | 2,823 (12%) |
|   HER2 enriched | 105 (6.5%) | 969 (4.1%) |
|   TNBC | 323 (20%) | 2,546 (11%) |
|   Unknown | 63 (3.9%) | 1,039 (4.4%) |
| Race and Ethnicity | | |
|   NH White | 903 (56%) | 15,248 (65%) |
|   NH Black | 592 (37%) | 6,622 (28%) |
|   Hispanic | 77 (4.8%) | 917 (3.9%) |
|   NH API | 33 (2.1%) | 608 (2.6%) |
|   NH Other | <15 (<0.1%) | 44 (0.2%) |
| Marital Status* | | |
|   Married | 840 (52%) | 13,115 (56%) |
|   Single (never married) | 279 (17%) | 3,269 (14%) |
|   Divorced | 238 (15%) | 3,242 (14%) |
|   Widowed | 187 (12%) | 2,886 (12%) |
|   Unknown | 62 (3.9%) | 927 (4.0%) |
| County | | |
|   Urban | 1,386 (86%) | 19,117 (82%) |
|   Rural | 220 (14%) | 4,322 (18%) |
| Treated at a CoC Facility | | |
|   Yes | 1,473 (92%) | 21,258 (91%) |
|   No | 133 (8.3%) | 2,181 (9.3%) |
| Any Medicare coverage | | |
| Yes | 899 (56%) | 11,847 (51%) |
| No | 707 (44%) | 11,592 (49%) |
| Any HMO coverage** | | |
|   Yes | 280 (17%) | 5,446 (23%) |
|   No | 1,326 (83%) | 17,993 (77%) |

Abbreviations: API: Asian or Pacific Islander, CoC: Commission on Cancer, GCR: Georgia Cancer Registry, ER: Estrogen receptor, HER2: human epidermal growth factor receptor 2, HMO: Health Maintenance Organization, NH: Non-Hispanic, TNBC: triple negative breast cancer.
*Married includes common law and domestic partnership, divorced includes separated
**Among patients with insurance through government programs

Table 2. Descriptive characteristics of 1,814 recurrent cases identified among female stage I–III breast cancer patients in GCR, diagnosed 2013 to 2017 with follow-up through 2019.

| Characteristic | N (%) |
| --- | --- |
| **Diagnosis Year** | |
| 2013 | 463 (26%) |
| 2014 | 445 (25%) |
| 2015 | 398 (22%) |
| 2016 | 306 (17%) |
| 2017 | 202 (11%) |
| **Age** | |
| 18-34 | 60 (3.3%) |
| 35-44 | 308 (17%) |
| 45-54 | 473 (26%) |
| 55-64 | 438 (24%) |
| 65-74 | 333 (18%) |
| 75-100 | 202 (11%) |
| **Stage** | |
| I | 389 (21%) |
| II | 802 (44%) |
| III | 623 (34%) |
| **Grade** | |
| 1 | 100 (5.5%) |
| 2 | 636 (35%) |
| 3+ | 1,013 (56%) |
| Unknown | 65 (3.6%) |
| **Tumor Subtype** | |
| Luminal A | 993 (55%) |
| Luminal B | 227 (13%) |
| HER2 enriched | 120 (6.6%) |
| TNBC | 390 (21%) |
| Unknown | 84 (4.6%) |
| **Race and ethnicity** | |
| NH White | 976 (54%) |
| NH Black | 663 (37%) |
| Hispanic | 94 (5.2%) |
| NH API | 36 (2.0%) |
| NH Other | 45 (2.5%) |
| **Marital Status*** | |
| Married | 930 (51%) |
| Single (never married) | 305 (17%) |
| Divorced | 315 (17%) |
| Widowed | 200 (11%) |
| Unknown | 64 (3.5%) |
| **County** | |
| Urban | 1,535 (85%) |
| Rural | 279 (15%) |

Abbreviations: API: Asian or Pacific Islander, GCR: Georgia Cancer Registry, HER2: human epidermal growth factor receptor 2, NH: Non-Hispanic, TNBC: triple negative breast cancer.
*Married includes common law and domestic partnership, divorced includes separated

Table 3. Multivariable adjusted HR and 95% CI for breast cancer recurrence among female stage I–III breast cancer patients in GCR, diagnosed 2013 to 2017 with follow-up through 2019.

| Characteristic | HR (95% CI) |
|---|---|
| **Age** | |
| 18-34 | 0.74 (0.45, 1.23) |
| 35-44 | 1.19 (0.77, 1.82) |
| 45-54 | 0.94 (0.65, 1.36) |
| 55-64 | Ref. |
| 65-74 | 0.93 (0.64, 1.34) |
| 75-100 | 1.32 (0.84, 2.08) |
| **Grade** | |
| 1 | Ref. |
| 2 | 1.59 (0.90, 2.82) |
| 3+ | 2.23 (1.26, 3.97) |
| Unknown | 1.88 (0.69, 5.10) |
| **Stage** | |
| I | Ref. |
| II | 1.93 (1.38, 2.68) |
| III | 4.1 (2.94, 5.72) |
| **Subtype** | |
| Luminal A | Ref. |
| Luminal B | 0.92 (0.63, 1.35) |
| HER2 enriched | 0.85 (0.55, 1.31) |
| TNBC | 1.25 (0.94, 1.67) |
| Unknown | 1.11 (0.60, 2.04) |
| **Race and Ethnicity** | |
| NH White | Ref. |
| NH Black | 0.94 (0.73, 1.21) |
| NH API | 1.03 (0.45, 2.39) |
| Hispanic | 1.35 (0.75, 2.44) |
| NH Other | 20.4 (5.94, 69.9) |
| **Marital Status*** | |
| Married | Ref. |
| Single (never married) | 1.00 (0.72, 1.39) |
| Divorced | 1.53 (1.04, 2.26) |
| Widowed | 1.23 (0.81, 1.87) |
| Unknown | 0.67 (0.44, 1.02) |
| **County** | |
| Urban | Ref. |
| Rural | 0.69 (0.50, 0.95) |

Abbreviations: API: Asian or Pacific Islander, CI: confidence interval, GCR: Georgia Cancer Registry, HER2: human epidermal growth factor receptor 2, HR: hazard ratio, NH: Non-Hispanic, TNBC: triple negative breast cancer.

*Married includes common law and domestic partnership, divorced includes separated

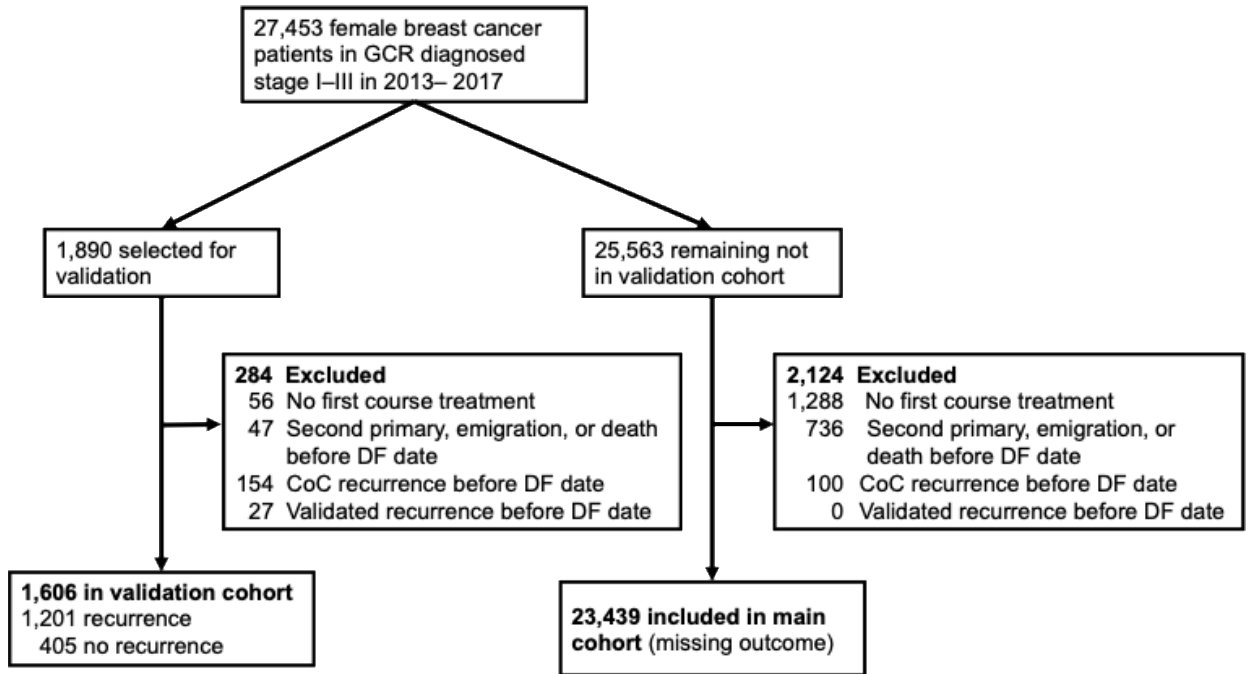

Figure 1. Study population selection diagram.

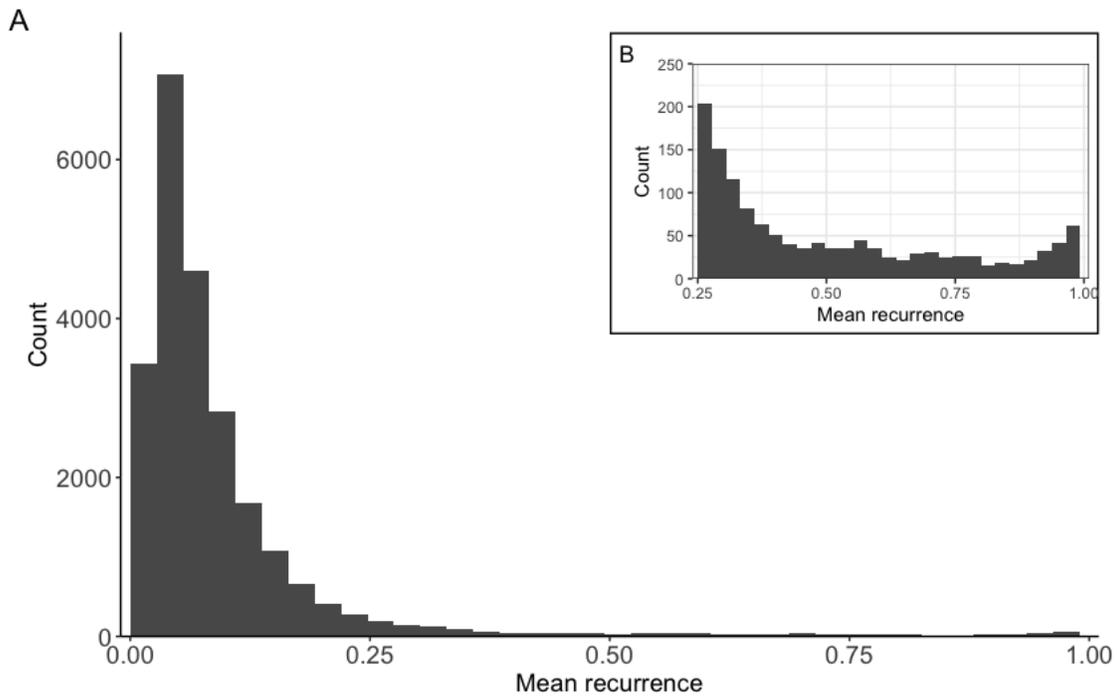

Figure 2. Histogram of mean imputed recurrence outcome probability for A) all patients in main cohort and B) subset with probability ≥ 0.25.